\documentclass[prl,twocolumn,nofootinbib]{revtex4-2}
\usepackage{graphicx}
\usepackage{color}
\usepackage{dcolumn}
\usepackage{bm}
\usepackage{amssymb}
\usepackage{footnote}
\usepackage{lipsum}
\usepackage{natbib}
\usepackage{wrapfig}
\usepackage{sidecap}%use to put caption beside figure
\renewcommand{\thefootnote}{\fnsymbol{footnote}}

\newcommand\blfootnote[1]{%
\begingroup
\renewcommand\thefootnote{}\footnote{#1}%
\addtocounter{footnote}{-1}%
\endgroup}
\begin{document}

\title{Quantum Hall phase in graphene engineered by interfacial charge coupling}

\author{Yaning Wang,$^{1,2,3,4,14*}$ Xiang Gao,$^{1,3*}$ Kaining Yang,$^{1,3*}$ Pingfan Gu,$^{5,6*}$ Xin Lu,$^{7}$ Shihao Zhang,$^{7,8}$ Yuchen Gao,$^{5,6}$ Naijie Ren,$^{1,3}$ Baojuan Dong,$^{1,3}$ Yuhang Jiang,$^{9}$ Kenji Watanabe,$^{10}$ Takashi Taniguchi,$^{11}$ Jun Kang,$^{12}$ Wenkai Lou,$^{13}$ Jinhai Mao,$^{9\dagger}$ Jianpeng Liu,$^{7,8\dagger}$ Yu Ye,$^{5,6\dagger}$ Zheng Vitto Han,$^{1,3,14\dagger}$ Kai Chang,$^{13\dagger}$ Jing Zhang,$^{1,3}$ Zhidong Zhang$^{2,4}$}

\affiliation{$^{1}$State Key Laboratory of Quantum Optics and Quantum Optics Devices, Institute of Opto-Electronics, Shanxi University, Taiyuan 030006, P. R. China}
\affiliation{$^{2}$Shenyang National Laboratory for Materials Science, Institute of Metal Research, Chinese Academy of Sciences, Shenyang 110016, China}
\affiliation{$^{3}$Collaborative Innovation Center of Extreme Optics, Shanxi University, Taiyuan 030006, P.R.China}
\affiliation{$^{4}$School of Material Science and Engineering, University of Science and Technology of China, Shenyang 110016, China}
\affiliation{$^{5,}$Collaborative Innovation Center of Quantum Matter, Beijing 100871, China}
\affiliation{$^{6}$State Key Lab for Mesoscopic Physics and Frontiers Science Center for Nano-Optoelectronics, School of Physics, Peking University, Beijing 100871, China}
\affiliation{$^{7}$School of Physical Science and Technology, ShanghaiTech University, 201210, Shanghai, China}
\affiliation{$^{8}$ShanghaiTech Laboratory for Topological Physics, ShanghaiTech University, 201210, Shanghai, China}
\affiliation{$^{9}$School of Physical Sciences and CAS Center for Excellence in Topological Quantum Computation, University of Chinese Academy of Sciences, Beijing, China}
\affiliation{$^{10}$Research Center for Functional Materials, National Institute for Materials Science, 1-1 Namiki, Tsukuba 305-0044, Japan}
\affiliation{$^{11}$International Center for Materials Nanoarchitectonics, National Institute for Materials Science,  1-1 Namiki, Tsukuba 305-0044, Japan}
\affiliation{$^{12}$Beijing Computational Science Research Center, Beijing 100193, China}
\affiliation{$^{13}$State Key Laboratory for Superlattices and Microstructures, Institute of Semiconductors, Chinese Academy of Sciences, P. O. Box 912, Beijing 100083, China}
\affiliation{$^{14}$Liaoning Academy of Materials, 280 Chuangxin Road, Shenyang 110167, China}

%\date{\today}
%\pacs{}

\maketitle
\blfootnote{\textup{*} These authors contribute equally.}

\blfootnote{$^\dagger$Corresponding to: jhmao@ucas.ac.cn, ye$\_$yu@pku.edu.cn, liujp@shanghaitech.edu.cn, kchang@semi.ac.cn, and vitto.han@gmail.com}

\textbf{Quantum Hall effect (QHE), the ground to construct modern conceptual electronic systems with emerging physics \cite{QHE_SC_Science_2016, Finkelstein_SA_2019, Linder_PRX_2012, Clarke_NC_2013, Mong_PRX_2014}, is often much influenced by the interplay between the host two-dimensional electron gases and the substrate, sometimes predicted to exhibit exotic topological states \cite{ZHQiao_PRL, YaoYugui_SCiRep_2015}. Yet the understanding of the underlying physics and the controllable engineering of this paradigm of interaction remain challenging. Here we demonstrate the observation of an unusual QHE, which differs markedly from the known picture, in graphene samples in contact with an anti-ferromagnetic insulator CrOCl equipped with dual gates. Two distinct QH phases are developed, with the Landau levels in monolayer graphene remaining intact at the conventional phase, but largely distorted for the interfacial-coupling phase. The latter QH phase even presents in the limit of zero magnetic field, with the consequential Landau quantization following a parabolic relation between the displacement field $D$ and the magnetic field $B$. This characteristic prevails up to 100 K in a sufficiently wide effective doping range from 0 to 10$^{13}$ cm$^{-2}$. Our findings thus open up new routes for manipulating the quantum electronic states, which may find applications in such as quantum metrology.}

 \bigskip

The QHE is found in a number of solid-state systems to demonstrate topologically protected dissipation-less edge channels with their transversal conductance quantized by $e^{2}/h$, with $e$ and $h$ being the elementary charge and the Planck constant, respectively \cite{1980_QHE_PRL, von_Klitzing_AR, Nat_Rev_Phys_2020, Laughlin_PRB_1981, Physics_Today_2003}. This peculiar behaviour has been crucial for such as the implementation of quantum-based resistance standards with extremely high precision and reproducibility \cite{LNE_NN_2015}. Among the few known systems manifesting QHE, graphene received special attention for its distinct band structure and the resulting N$^{th}$ Landau level (LL) at the energy of $\varepsilon_\mathrm{LL}(N)=v_{F}\mathrm{sgn}(N)\sqrt{2e\hbar B|N|}$ under a fixed magnetic field $B$. Due to the linear transformation of energy and the square root of carrier density $n$ \cite{RT_QHE_Science, Yuanbo_Nature, Novoselov_Nature_2005}, the Landau quantization of conventional graphene in the parameter space of $B$ and $n$ is defined as the famed Landau fan diagram, with all LLs linearly extrapolated to the charge neutrality point \cite{Yuanbo_Nature, DGG_NC_2015, Novoselov_Nature_2005}.

   \begin{figure*}[ht!]
   \includegraphics[width=0.89\linewidth]{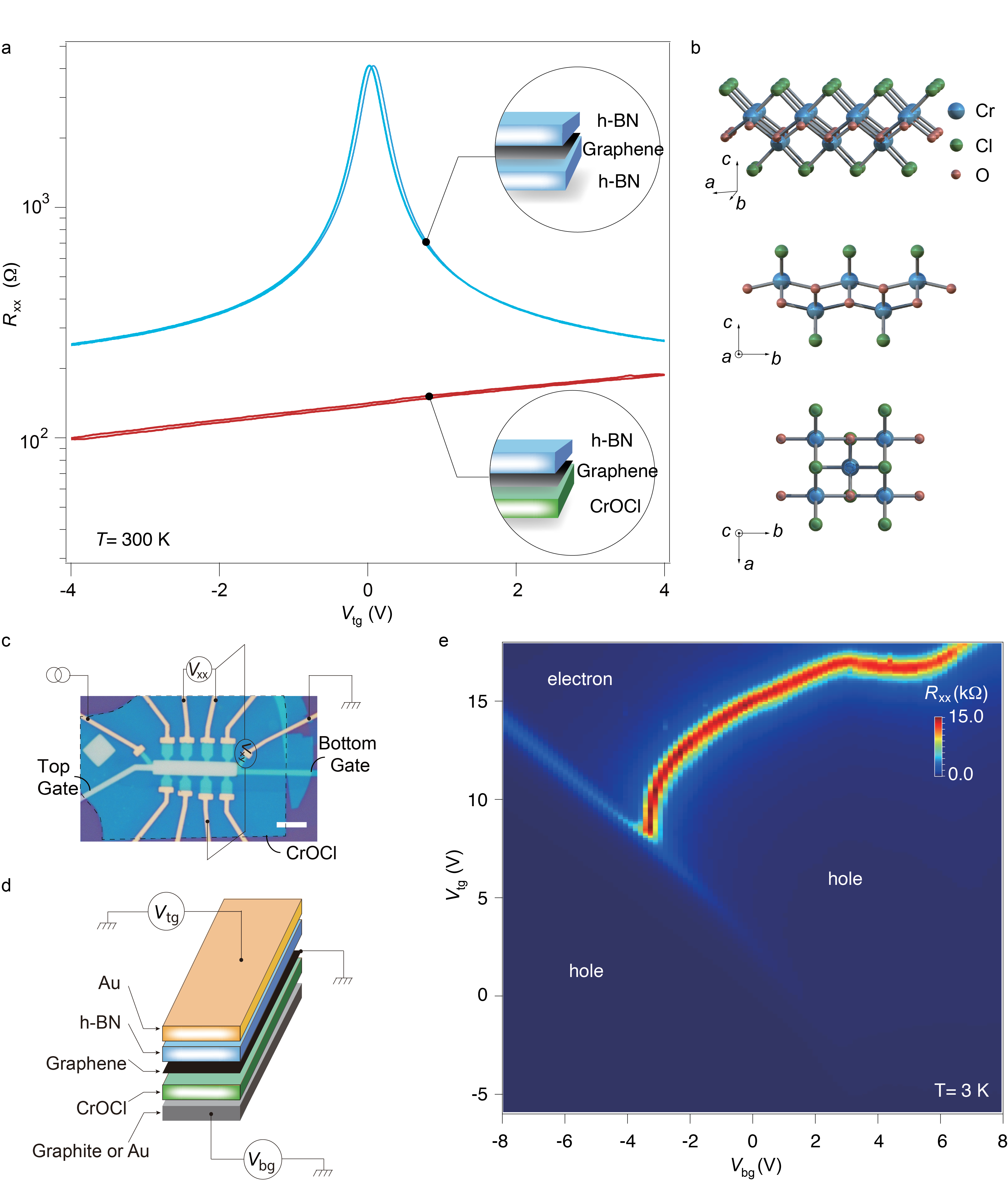}
   \caption{\textbf{Characterization of CrOCl supported graphene.} (a) Field effect curves of graphene encapsulated with h-BN and/or CrOCl, with each configuration illustrated in the schematic cartoon in the inset. (b) Schematics of the crystallographic structure of CrOCl. (c) Optical micrograph image of a typical h-BN/graphene/CrOCl sample, with its cartoon illustration shown in (d). Scale bar is 5 $\mu$m. (e) Color map of a dual gate scan of field effect in a typical sample, measured at a temperature of $T$= 3 K and at magnetic filed $B$=0.
   }
   \end{figure*}

Interfacial coupling is known to be an important factor to affect the QHE in graphene, usually by two different means: charge impurities that cause reduced mobility leading to wider QH plateaux at some circumstances \cite{LNE_NN_2015}, and charge transfer that to some extent shifts the effective doping \cite{Sci_Rep_2016, Photo_Doped_QHE, AM_PhotoChemDoping, Kubatkin_APL_2014, Berger_JPCB_2004, InSe_PRL_Plateau_2017}. Recent theories predict that the interplay between an antiferromagnetic insulator substrate and a graphene layer can give rise to topological quantum ground states such as quantum anomalous Hall phases \cite{ZHQiao_PRL, YaoYugui_SCiRep_2015, Takenaka_PRB2019}. Experimentally, interfacial coupling from such as RuCl$_{3}$ to graphene is indeed spotted, with  strong charge transfer, which is sometimes possibly coupled to the magnetism \cite{RuCl3_PRB} while sometimes not fully evidenced so \cite{RuCl3_NanoLett}. To date, understanding of the underlying physics and controllable engineering of the interfacial coupling between the specific insulator substrate and graphene, especially in the QHE regime remains largely unknown.

In this work, we investigate the case of monolayer graphene interfaced to CrOCl, an antiferromagnetic insulator with Cl atoms as the ending bonds at its surface. By examining multiple configurations of graphene encapsulated with h-BN and/or CrOCl, we mapped out the peculiar interfacial coupling between the carbon honeycomb lattice and CrOCl in the parameter space of temperature $T$, total gate doping $n\mathrm{_{tot}}$, magnetic field $B$, and electrical field $D$. At low temperatures, where the CrOCl bulk is totally insulating, strong interfacial coupling (SIC) is found at positive gate voltages. At finite magnetic fields, it leads to a gate-tunable crossover from fan-like to cascades-like Landau quantization. In the regime of positive total electron doping, a QHE phase with parabolic dependence between $B$ and $D$ is obtained, with the Landau quantization reaching the zero magnetic field limit. This way allows one to effectively engineer the QHE in this hybrid system, with a $\nu$ =$\pm 2$ plateau starting from as low as sub 100 mT and prevails up to 100 K in a wide doping range from 0 to 10$^{13}$ cm$^{-2}$. Our results open up new routes for manipulating the quantum electronic states via SIC, which may lead to evolutional applications in such as quantum metrology in the $\textit{Syst$\grave{e}$me International d'unit$\acute{e}$s}$.

%cascades-like Landau quantization at positive filling fractions, which largely distorted from the original fan-like pattern in conventional graphene. Notably, the Landau Levels in monolayer graphene remain intact at negative filling fractions. This

\bigskip
%\section{Results}
\textbf{Characterizations of graphene/CrOCl heterostructures.} Monolayered graphene as well as thin CrOCl flakes and encapsulating hexagonal boron nitride (h-BN) flakes were exfoliated from high-quality bulk crystals and stacked in ambient condition using the dry transfer method.\cite{Lei_Science} The vertically assembled van der Waals heterostructures were then patterned into Hall bars with their electrodes edge-contacted. As seen in Fig. 1a, the field-effect curve of h-BN/graphene/CrOCl samples (red curve) differs significantly with the conventional h-BN/graphene/h-BN ones (blue curve), with the resistive Dirac peak disappearing and the conductivity, in general, is enhanced compared with the former structure but with degraded gate tunability (art views of each configurations are illustrated in inset of Fig. 1a). Figure 1b shows the crystal structure of CrOCl, which consists of a Cr-O double-layer sandwiched between the Cl atom layers stacked along the $c$-axis \cite{Thesis_Reuvekamp_2014}.

We first started with single gated devices and found that an SIC is taking place to affect the real doping in graphene that exhibits significant discrepancy from the doping expected from conventional gate dielectric, as shown in Supplementary Figures 1-6. Fig.1c shows the optical micrograph image of a typical h-BN/graphene/CrOCl heterostructure with top and bottom gates, with a cartoon illustrating the layered structure of such typical devices (Fig. 1d). An example of dual-gate mapping of resistance obtained at a temperature of $T$=3 K is given in Fig.1e. Three notable regions are seen, each separated by a resistive peak and marked as either hole or electron doping, which is obtained via the analysis of the measurements at high magnetic fields, as will be discussed in the coming sections. To further elucidate the specific state of the SIC in the current hybrid system, we fabricated devices equipped with dual gates so that an effective displacement field $D_\mathrm{eff}=(C_\mathrm{tg}V_\mathrm{tg}-C_\mathrm{bg}V_\mathrm{bg})/2\epsilon_{0} - D_{0}$ can be applied to between the top and bottom gates, and at the same time the total carrier $n_\mathrm{tot}=(C_\mathrm{tg}V_\mathrm{tg}+C_\mathrm{bg}V_\mathrm{bg})/e-n_{0}$ induced by the dielectrics from top and bottom gates can be tuned, as previously used in multiple-layered graphene devices \cite{FengWang_Nature_BLG, Maher_Science}. Here, $C_\mathrm{tg}$ and $C_\mathrm{bg}$ are the top and bottom gate capacitances per area, respectively. And $V_\mathrm{tg}$ and $V_\mathrm{bg}$ are the top and bottom gate voltages, respectively. $n_{0}$ and $D_{0}$ are residual doping and residual displacement field, respectively. Notice that the real doping in graphene $n_\mathrm{graphene}$  that can be probed by transport measurement is affected by the interfacial states of CrOCl (see Supplementary Note 1), which can be different from $n_\mathrm{tot}$ in the interfacial coupling phase, as will be discussed in the coming text. Examples of plotting the dual gated map of channel resistance into the space of $D_\mathrm{eff}-n_\mathrm{tot}$ are given in Supplementary Figures 7-12.

   \begin{figure*}[ht!]
   \includegraphics[width=0.98\linewidth]{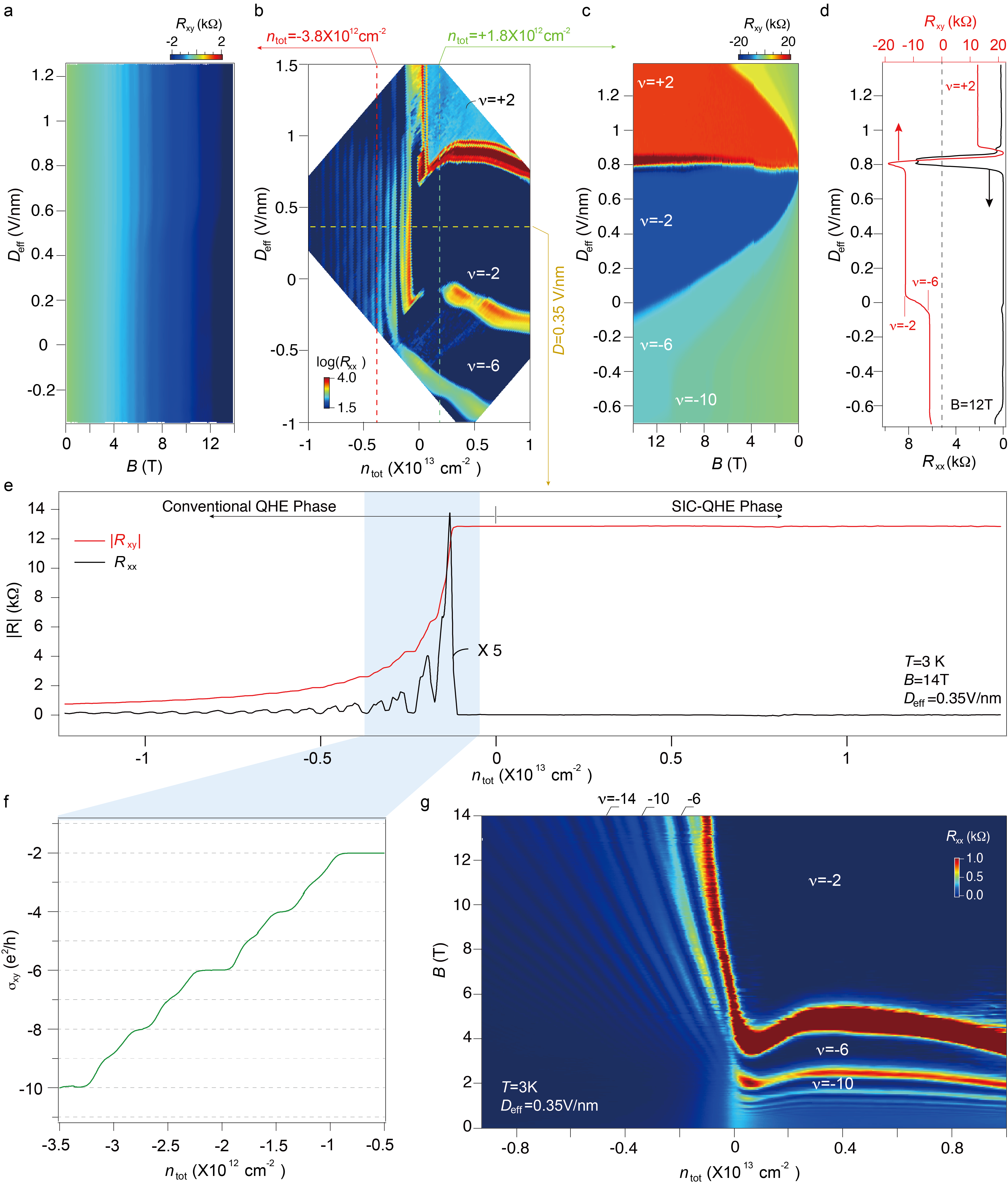}
   \caption{\textbf{Gate tunable SIC in the quantum Hall regime.} (a) and (c) are color maps of $R_{xy}$ as a function of magnetic field $B$, recorded along the red and green dashed lines in (b), respectively. (b) $R_{xx}$ in the parameter space of $D_\mathrm{eff}$ and $n_\mathrm{tot}$, measured at 14 T and 3 K. (d) Line profile of $R_{xx}$ and $R_{xy}$ at $B$=12 T in the color map in (c). (e) Line profile of $R_{xx}$ and $R_{xy}$ at $D_\mathrm{eff}$=0.35 V/nm (along the yellow dashed line) in the color map in (b), with the zoomed-in $\sigma_{xy}$ in the smaller range of hole doping shown in (f). (g) Gate tunable cross over from fan-like to cascades-like Landau quantization at $D_\mathrm{eff}$=0.35 V/nm.}

   \label{fig:fig2}
   \end{figure*}

\bigskip
\textbf{Gate tuned SIC in the quantum Hall regime.} Figure 2a shows a magnetic field scan of $R_\mathrm{xy}$ along fixed carrier density at the hole side with $n_{\mathrm{tot}}\sim -3.8\times 10^{12} \mathrm{cm}^{-2}$ (red dashed line in Fig. 2b, a mapping of channel resistance of Device-S16 in the $D_\mathrm{eff}-n_\mathrm{tot}$ space). Little $D_\mathrm{eff}$-dependence of filling fraction ($i.e.$, LLs) is seen. This is a typical behavior of monolayer graphene, as there is no $z$ dimension and thus the displacement field does not affect the LLs in it. Strikingly, shown in Fig. 2c, magnetic field scan of transverse resistance $R_\mathrm{xy}$ along $n_{\mathrm{tot}}\sim +1.8\times 10^{12} \mathrm{cm}^{-2}$ (green dashed line in Fig. 2b) exhibits drastically different patterns as compared to that in Fig. 2a. As discussed in the previous sections, with a single gate, only hole doping can be seen in graphene/CrOCl heterostructures (Supplementary Figure 4-5), and more details of carrier types in the dual gated devices can be found in Supplementary Figure 13. Here, at the fixed $n_{\mathrm{tot}}$ applied in Fig. 2c, dual gate sweep allows one to reach the electron side at $D_\mathrm{eff} \sim$0.8 V/nm, as clearly indicated by the line profiles of both $R_\mathrm{xx}$ and $R_\mathrm{xy}$ at 12 T in Fig. 2d, where a well-developed quantum Hall plateau of $\nu=+2$ can be seen. In this regime of $n_\mathrm{tot}$ and $D_\mathrm{eff}$ (we call it SIC-QHE phase), $R_\mathrm{xy}$ is quantized in an extremely wide parameter space. For example, at $B$=14 T, filling fraction $\nu=\pm 2$ is found in the doping range of $n_{\mathrm{tot}}$ from 0 to 10$^{13}$ cm$^{-2}$, and a displacement difference $\delta D$ of over $\sim$ 2 V/nm, which converts into a very large range of gate voltages. Notably, the quantized $R_\mathrm{xy}$ can be found in very low $B$, reaching the zero magnetic field limit, which will be discussed in more details in the coming sections.

Fig. 2e shows the line profiles of $R_\mathrm{xx}$ and $R_\mathrm{xy}$ at a fixed displacement $D_\mathrm{eff}$ = 0.35 V/nm (along the yellow dashed line in Fig. 2b) at $B$=14 T and $T$=3 K. It is seen that on the hole side (noted as conventional QHE phase) in the curves, Landau quantizations are in agreement with that observed in conventional monolayered graphene \cite{Novoselov_Nature_2005,Yuanbo_Nature}. Fig.2f illustrates a zoomed-in view of transverse conductivity $\sigma_\mathrm{xy}$ at $n_{\mathrm{tot}}$ between -0.5 $\times$ 10$^{12}$ cm$^{-2}$ and -3.5 $\times$ 10$^{12}$ cm$^{-2}$, indicating full degeneracy lifting with quantized plateaux at each integer filling fractions from $\nu$=-2 to -10. This speaks the high mobility of the graphene in the conventional QHE phase in the graphene/CrOCl heterostructure. Indeed, by fitting of the hole-side part of effect curve at zero magnetic field (Supplementary Figure 14), hole carrier mobility is estimated to be about 10$^{4}$ cm$^{2}$V$^{-1}$s$^{-1}$, which is comparable to the value in high-quality monolayer graphene reported elsewhere \cite{Lei_Science}. 

\begin{figure*}[ht!]
 \includegraphics[width=0.93\linewidth]{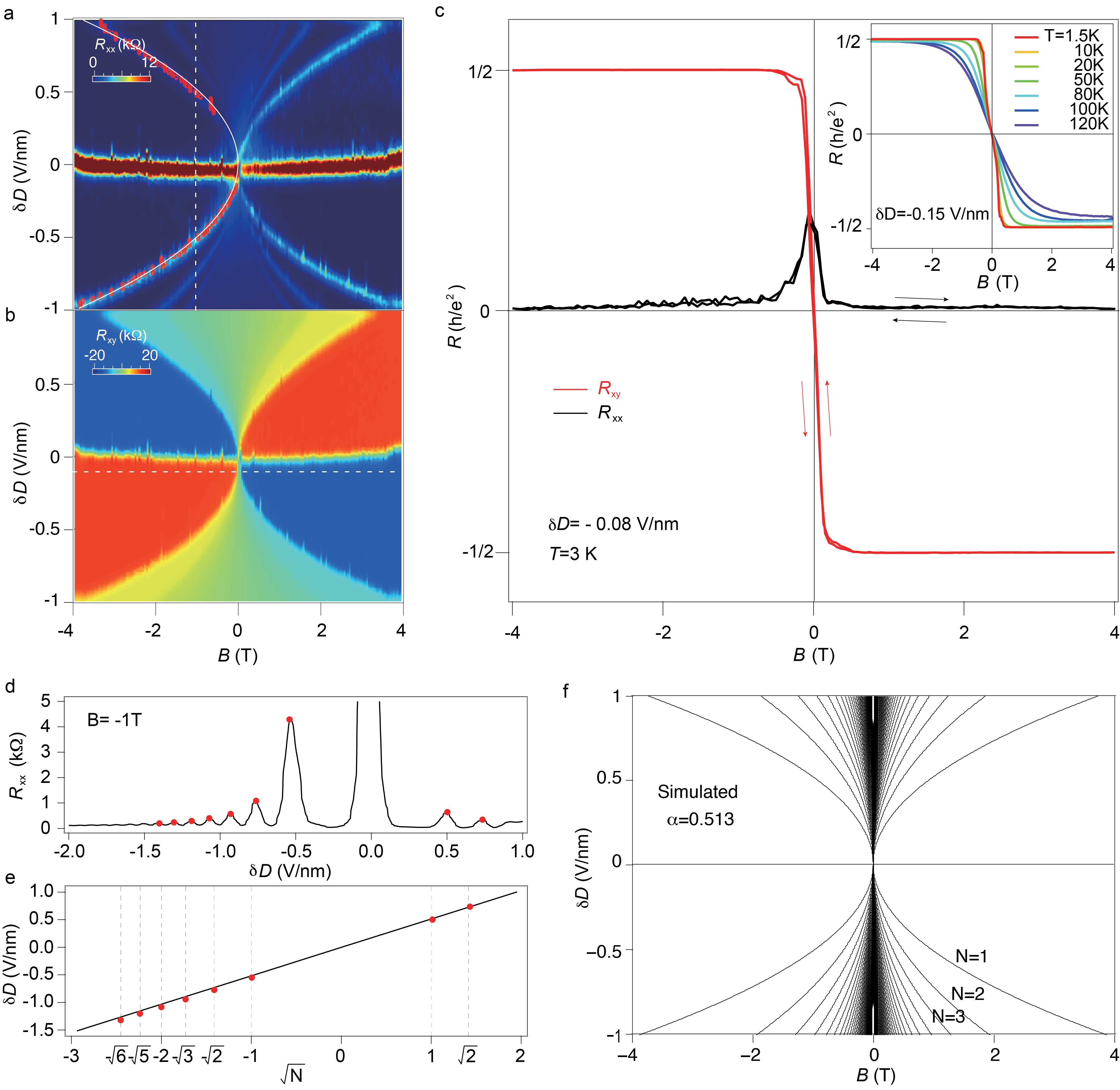}
 \caption{\textbf{Characteristics of the SIC-QHE phase in graphene/CrOCl heterostructures.} (a) $R_\mathrm{xx}$ and (b) $R_\mathrm{xy}$ plotted in the parameter space of $\delta D$ and $B$.   (c) Line profiles of $R_\mathrm{xx}$ and $R_\mathrm{xy}$ at $\delta D$= -0.08 V/nm. Inset shows temperature dependence of another typical sample at $\delta D$= -0.15 V/nm. (d) Line profile of $R_\mathrm{xx}$ in Fig. 3a at $B$= -1 T (indicated by the vertical white dashed line). Red dots are resistive peaks picked by each maximum. (e) Dependence of $\delta D$ and $\sqrt{N}$. Black solid line is a linear fit. (f) Parabolic dependence of $\delta D= \alpha \sqrt{B\cdot |N|}$, plotted with $\alpha$=0.513, and $|N|<200$. }
 
 \label{fig:fig3}
 \end{figure*}

On the electron side of the line profiles of $R_\mathrm{xx}$ and $R_\mathrm{xy}$ in Fig. 2e, a SIC-QHE phase dominates, as the quantum Hall plateau of $\nu$=-2 extends through out the whole gate range, up to +1.5 $\times$ 10$^{13}$ cm$^{-2}$. By varying the magnetic fields along fixed $D$=0.35 V/nm, one obtains a color map in the parameter space of $B$ and $n_{\mathrm{tot}}$, shown in Fig. 2g. It can be seen that the gate tunable SIC leads to a change of Landau quantization from the well-known fan-like behavior to a cascades-like one, as the system undergoes a phase transition from conventional QHE to SIC-QHE. To verify the $n_{\mathrm{graphene}}$ as compared to $n_{\mathrm{tot}}$ in the sample in the low field limit ($i.e., B <$ 0.5 T before the quantum oscillation starts) in Fig. 2g, we extracted $n_{\mathrm{eff}}$ from Hall resistance at fixed $D$ (0.35 V/nm ), which shows a slope of $\sim $1 with $n_{\mathrm{tot}}$ at the conventional phase, but strongly departures at positive $n_{\mathrm{tot}}$, as shown in Supplementary Figure 15. Moreover, to have a global picture of the major features described above, the color-maps shown in Fig.2 are re-plotted in a 3D presentation, as shown in Supplementary Figure 16. All these observations are reproducible in multiple samples (Supplementary Figure 17-18), and also confirmed in samples fabricated in a glove box, ruling out possible strong moisture or air adsorption in the interface of graphene/CrOCl heterostructures (Supplementary Figure 19). 

%, with the resistivity of the device generally higher at $\theta_\textrm{t}$/$\theta_\textrm{b}$ = 0$^{\circ}$/0$^{\circ}$, as compared to that of 0$^{\circ}$/60$^{\circ}$ configurations

%\section{Discussion}
\bigskip

\textbf{Characteristics of the SIC-QHE phase.} Central result of this letter is the observation of a SIC-QHE phase, where Landau quantizations seem to be ``pinned'' at a fixed filling fraction at a certain hole doping and at a fixed $D_\mathrm{eff}$, such as shown in Fig. 2g. A trivial explanation for this would be that the coupling at the graphene-CrOCl interface has accumulated holes that screen and cancel out the positive gate voltages applied, leading to a failure of electron injection and thus the total effective doping remains pinned. However, the most striking feature in this work, as shown in Fig. 2b-c, is that the displacement field $D_\mathrm{eff}$ totally shuffles the Landau quantization (and hence the $n_{\mathrm{graphene}}$), which rules out the ``charge pinning'' or ``gate screening'' picture, as the latter should be $D$-independent just like in the conventional QHE phase (such as Fig. 2a). Moreover, the Landau quantization seems to be approaching the $B$=0 limit in the SIC-QHE phase, as shown in Fig. 2c.

\begin{figure*}[ht!]
 \includegraphics[width=0.90\linewidth]{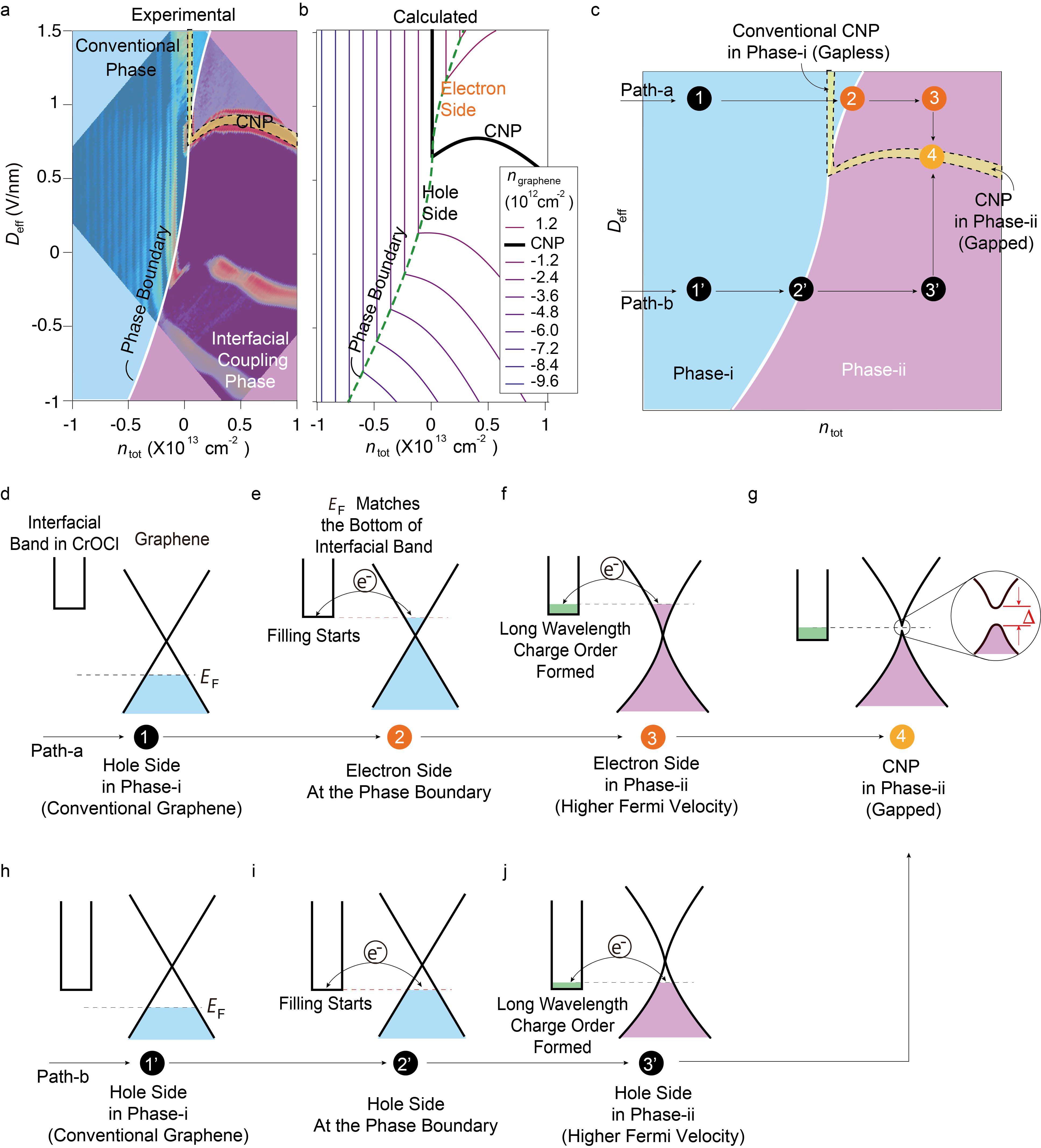}
 \caption{\textbf{QH phase diagram in the $D_\mathrm{eff}$-$n_\mathrm{tot}$ space and the transition processes between phases.} (a)-(b) Experimental and calculated phase diagram in the $D_\mathrm{eff}$-$n_\mathrm{tot}$ space, with the phase boundary highlighted. Iso-doping lines with the calculated $n_\mathrm{graphene}$ are indicated by solid lines in (b). (c) Schematic of the typical phase diagram with two paths of transition processes noted by arrows. (d)-(g) and (h)-(j) are schematics of the band diagrams for Path-a and Path-b illustrated in (c).}
 
 \label{fig:fig4}
 \end{figure*}

To further clarify this perplexing scenario, we carried out a zoomed-in scan of the low magnetic field part of Fig. 2c. We define the displacement field where carrier type switches from holes to electrons as $D_\mathrm{Neutral}$, thus the $D$ axis can be renormalized as $\delta D= D -D_\mathrm{Neutral}$. As shown in Fig. 3a-b ($R_\mathrm{xx}$ in Fig. 3a, and $R_\mathrm{xy}$ in Fig. 3b), $\delta D$ was scanned in the range of -1 to +1 V/nm, while the magnetic field was scanned from -4 to +4 T. Wide Landau plateaux are seen in Fig. 3b, with the quantized regions indeed touching the $B$=0 T line, while a tiny width still exists at the vicinity of zero magnetic field, due to the signal swap from positive to negative sign. The $D$-$B$ relation of LLs observed here is distinct from those found in other multilayered graphene systems \cite{Weitz_Science_2010, Javier_PRL2012, Lau_PNAS2019}.

By bringing $\delta D$ infinitely close to  $\delta D$=0, one expects to have a Landau quantization at $\nu$= $\pm$2 at the zero magnetic field limit. Due to experimental noises, we take the $\delta D$= -0.08 V/nm here (indicated by the white dashed line in Fig. 3b), and plotted both $R_\mathrm{xx}$ and $R_\mathrm{xy}$ in Fig. 3c. The curves indeed show well quantized plateaux at $R_\mathrm{xy}$=$\pm$0.5 $\mathrm{h}/e^{2}$ starting from $B$ as low as sub 100 mT, with $R_\mathrm{xx}$ showing near zero reminiscent values at each plateau. Although quantum anomalous Hall effect (QAHE) or Chern insulator is claimed in suspended bilayer graphene and moir$\mathrm{\acute{e}}$ heterostructures such as aligned ABC-stacked trilayer graphene/h-BN,\cite{BLG_QAHE_Weitz, Young_Science, FengWang_Nature, Mak_Nature_intertwined} we found that our system of h-BN/monolayer-graphene/CrOCl heterostructure seems to be topologically trivial when magnetic field is completely absent, and the observed $R_\mathrm{xy}$ quantization at very low magnetic field is still in the regime of quantum Hall states, since the quantization of $\nu$ = $\pm 2$ is inherited from the Dirac electrons that are dominating in transport in the system, and no magnetic hysteresis ($i.e.$, the coercive field) is seen in the trace-retrace loop of magnetic scan in our system (indicated by red and black arrows in Fig.3c, and the zoomed-in scan in Supplementary Figure 20). More discussion can be seen in Supplementary Note 2. Trivial effect of gate leakage is ruled out, and multiple samples are tested to a maximum temperature before gate leakage takes place, shown in Supplementary Figure 21-23. Notably, this robust SIC-QHE phase in the graphene/CrOCl heterostructure prevails at much higher temperatures (inset of Fig. 3c) as compared to the QAHE systems so far reported, while the latter often requires temperatures of sub 2 K or even dilution fridge temperatures \cite{Young_Science, MnBiTe_Science_2020, CuiZu_2013}. It also requires very relaxed experimental conditions as compared to such as doped (Hg,Mn)Te topological insulators quantum well, which is recently reported to show QH state below 50 mT at 20 mK \cite{Molenkamp_SA2020}. It thus allows future applications in quantum metrology in the $\textit{Syst$\grave{e}$me International d'unit$\acute{e}$s}$ \cite{Lafont_NC_2015, LNE_NN_2015}.

\begin{figure*}[ht!]
 \includegraphics[width=0.8\linewidth]{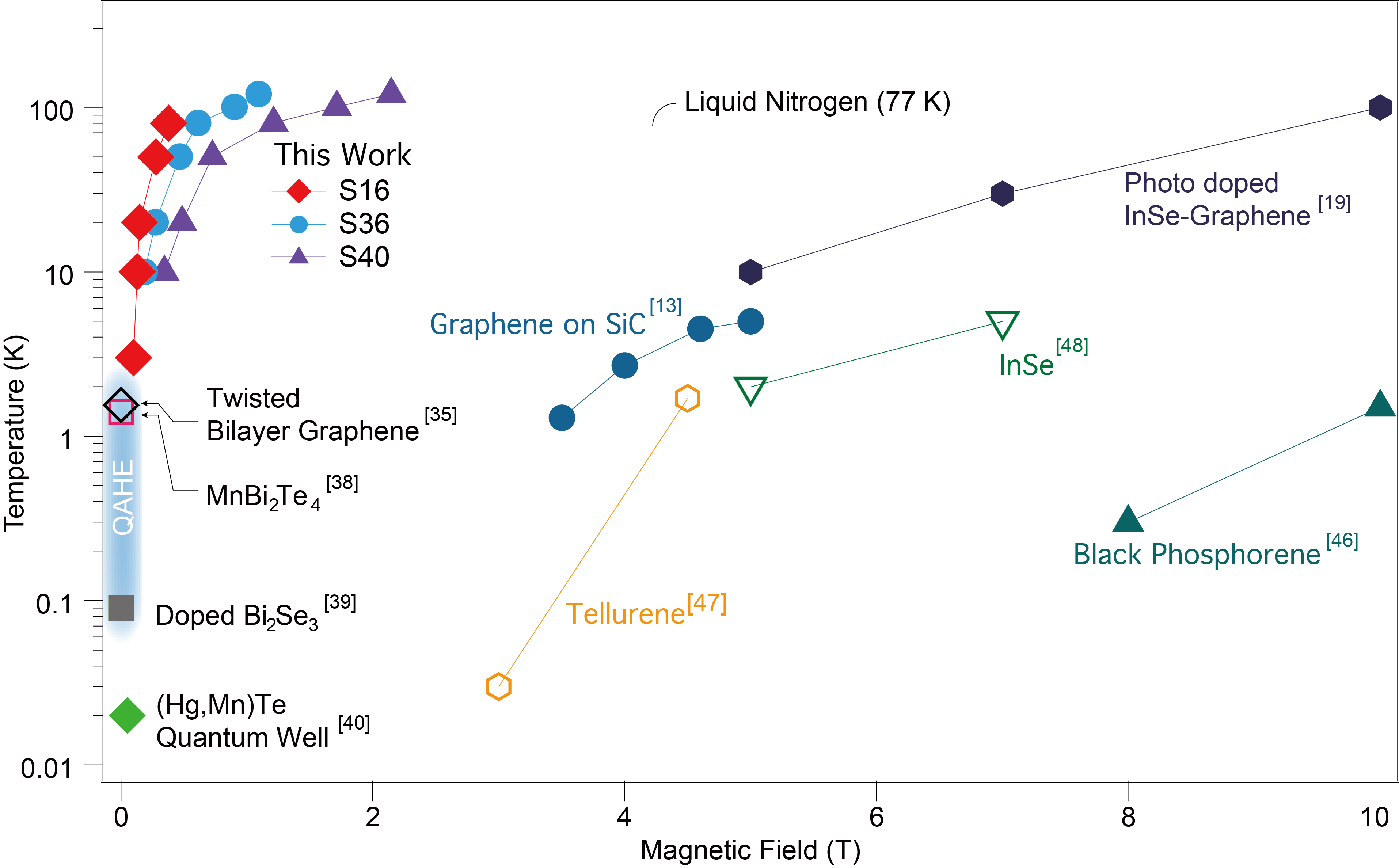}
 \caption{\textbf{Perspectives of the SIC-QHE.} The diagram summarizes magnetic fields (below 10 T) and temperatures for realizing quantized Hall conductance in several typical systems reported recently. Data from three devices (S16, S36, and S40) in this work is included, as indicated by the legend in the figure.}
 
 \label{fig:fig5}
 \end{figure*}
 
By extracting a line profile of $R_\mathrm{xx}$ in Fig. 3a at $B$= -1 T (indicated by the vertical white dashed line), resistive peaks can be found at each LLs, as indicated by the red dots in Fig. 3d. It is found that the $\delta D$ values at each resistive peak are in linear dependence with $\sqrt{N}$, with $N$ the $N^{th}$ LL, shown in Fig. 3e. This is a typical Landau quantization energy dependence in monolayered graphene. Indeed, $\delta D$-$B$ relation can be fitted using a parabolic curve as $\delta D= \alpha \sqrt{B\cdot |N|}$. Peaks of $R_\mathrm{xx}$ of the first LL in Fig.3a are picked as red circles, and fitted with a white solid parabolic curve, with $\alpha$=0.513. The first 200 LLs are then plotted in Fig. 3f, well simulating the experimental $\delta D$ data. This indicates that $\delta D$ linearly tunes the chemical potential of the Landau levels of graphene, which stimulates us to propose a possible mechanism in explaining the SIC-QHE to be explained in the following section.

Interestingly, the observed SIC-QHE phase seems to have no connection to the anti-ferromagnetic nature of CrOCl itself, as the N$\mathrm{\acute{e}}$el temperature of it is only $\sim$13 K, way lower than the upper bound temperature for the SIC-QHE phase. Recently, charge transfer phenomena have been also found in such as RuCl$_{3}$/graphene systems, which give rise to a shifting of Raman signal or plasmonic characteristics in graphene, and seem to be often irrelevant to the magnetic order of the Cl-based substrates\cite{RuCl3_Raman_NanoLett2020, Basov_Plasmon_NanoLett2020}. Those other Cl-based compounds are reported to lead to inferior mobilities in the resulted hybrid systems, and are not suitable for QHE studies \cite{RuCl3_PRB, RuCl3_NanoLett}. In addition, we noticed that, a sister compound of CrOCl, FeOCl is far less stable, and could not be used to check the universality of the findings in this work (Supplementary Figure 24).

%\bigskip
\textbf{Possible mechanism of the SIC-QHE at zero field limit.} We now replot in Fig. 4a the $R_\mathrm{xx}$ of Device-16 in the $D_\mathrm{eff}-n_\mathrm{tot}$ space at 12 T with falsed color that separates the boundary between the conventional and SIC phases. And the LLs naturally denote each iso-doping line associated to the fixed filling fraction $\nu$, as defined by $\nu=hn_\mathrm{graphene}/eB$. Two key features are seen in Fig. 3a. First, the charge neutrality point (CNP) is bent as the system enters from conventional phase into the SIC-phase. Second, the spacing between each iso-doping lines increases as the system enters deeper into the SIC phase. To explain them, we propose an electro-static model in Supplementary Note 1. An interfacial band with considerable density of states near the CNP of graphene is introduced at the surface of CrOCl with a distance of $d_{2}$ below the graphene layer. While top and bottom gates are located at distances of $d_{1}$ and $d_{3}$, respectively, as illustrated in Supplementary Figure 25a-b. By evaluating the model, we found that the above mentioned two major features can be well reproduced, as shown in the phase diagram in Fig. 3b. Nevertheless, one has to introduce in the simplified model two assumptions, including a band structure reconstruction with an enhanced Fermi velocity once the Fermi level of graphene touches the interfacial band, and also that the interfacial band exhibits no contribution to transport, as discussed in Supplementary Figures 25-28.

To elucidate the above simplified electro-static model, we performed DFT calculations considering systems with both monolayer and bilayer CrOCl supported graphene (Supplementary Figure 29-31 in Supplementary Note 2). It is seen that, in the bilayered CrOCl model and at certain vertical electric fields, the interfacial band from the top layer of CrOCl (mainly from the top Cr atoms, see Supplementary Figures 32-34) starts to overlap with the Dirac point of graphene, this is the event that charge transfer (i.e., filling of electrons in the interfacial band of CrOCl) from graphene to the interfacial band is triggered. Our DFT calculations suggest that long-wavelength localized charge order (Wigner crystal in our case, since the dimensionless Wigner-Seitz radii are estimated to exceed the critical value of 31 for 2D electrons,\cite{Wigner_radii} shown in Supplementary Table 1) is formed in the interfacial band of Cr $3d$ orbital in the top layer of CrOCl. It self-consistently explains that, once filled with electrons, the interfacial band undergoes Wigner instability and does not contribute to transport, but provides a superlattice of Coulomb potential for the graphene resting on top of CrOCl. When systematically considering the interplay between generic long-range Coulomb lattices and the Dirac electrons, our separate theoretical work suggests that such $e-e$ interaction in graphene indeed enhances the Fermi velocity dramatically and in the meantime opens a gap at the CNP.\cite{JPLiu_arXiv}

Based on the above analysis, we can then plot a schematic phase diagram in Fig. 3c, where the conventional and SIC phases are denoted as Phase-i and Phase-ii for simplicity. Two different paths are listed, to illustrate the doping processes of our system. In Path-a, graphene starts in a hole-doped conventional state (State-1 in Fig. 3d). It crosses the CNP, becomes electron-doped, and approaches the phase boundary where the Fermi level of graphene touches the lowest energy of the interfacial band in the top layer of CrOCl (State-2 in Fig. 3e), it thus triggers the electron-filling event (solid white phase boundary line in Fig. 3c) in the interfacial band. Upon entering in Phase-ii, the electrons filled in the interfacial band forms a Wigner crystal, which exerts a long-wavelength Coulomb potential to the Dirac electrons in graphene in such a way to renormalize the non-interacting band structures, thus $e-e$ interaction would play a more important role. Consequently, the Fermi velocity is significantly enhanced (sharpening of the Dirac cone in the illustration) driven by e-e interactions in graphene (State-3 in Fig. 3f). The system enters in Phase-ii, and remains electron-doped. Furthermore, when decreasing $D_\mathrm{eff}$ from State-3 to State-4, the Fermi level in graphene reaches its CNP, where an interaction-driven gap is seen (as supported by the extraction of thermal activation gap in Supplementary Figure 35). Upon further decreasing $D_\mathrm{eff}$, the system becomes hole-doped again. Similar process can be interpreted for Path-a, in Fig. 3h-j.

Experimentally, by fitting the Shubnikov-de Haas oscillations from various temperatures at dopings in Phase-i and Phase-ii (Supplementary Figure 36), the cyclotron mass m$^{*}$ in Phase-i is estimated to be comparable to that in ``ordinary'' monolayer graphene, but 3-5 times larger than that in Phase-ii. It thus yields a few times higher Fermi velocity of graphene in Phase-ii, in agreement with the conjectures in our simplified electro-static model and DFT calculations. Thus in this regime the cyclotron gap of first LL $\Delta=v_{F}\sqrt{2\hbar eB}$ will be in the order of about 100 meV at 0.1 T, which qualitatively explains the quantization at very low magnetic field. We emphasize that further probes such as infra-red transmission would be helpful in directly verifying the cyclotron gap estimated in the current system in this regard.

%Possibility of emerging topologically protected symmetries in the electronic band structures can be ruled out in two ways. First, the $D-B$ scan follows the $D\sim \sqrt{B|N|}$ relation, which is a consequence of LLs crossing the Fermi level at fixed $D$ and $n_{\mathrm{tot}}$, indicating that the quantization of $\nu$ = $\pm 2$ is inherited from the Dirac electrons that are still dominating in transport in the system. Second, no signature of topological physics is found in the current system via density functional theory calculations. 

A robust quantum Hall state with ultra-low magnetic fields at relaxed experimental conditions can be crucial for future constructions of topological superconductivity as well as quantum information processing, which has long thought to be only possible in QAHE systems. It unambiguously tells that the interfacial coupling, in terms of engineering the quantum electronic states, is a powerful technique that we may have overlooked thus far. For comparison, Fig. 5 shows a diagram summarizing magnetic fields and temperatures required for realizing quantized Hall conductance in typical different QHE or QAHE systems \cite{Young_Science, MnBiTe_Science_2020, CuiZu_2013, Molenkamp_SA2020, LNE_NN_2015, Photo_Doped_QHE, BP_QHE, Tellurene, InSe_QHE}, reported recently.

%\section{Conclusion}

\bigskip

In conclusion, we have demonstrated a hybrid system of graphene/CrOCl, in which an exotic QHE phase was observed thanks to the peculiar gate tunable interfacial coupling. At finite magnetic fields and constant $D_\mathrm{eff}$, a crossover from fan-like to cascades-like Landau quantization was seen. And in the $D$-$B$ space, unlike the conventional $D$-independent ones, LLs in the SIC-QHE phase exhibit parabolic dependence between $B$ and $D_\mathrm{eff}$ in a wide effective doping range from 0 to 10$^{13}$ cm$^{-2}$, with the Landau quantization of $\nu$ =$\pm 2$ plateau starting from as low as sub 100 mT below 10 K, and remains quantized at $\sim$ 350 mT at liquid nitrogen temperature. Our theoretical analysis self-consistently attribute the physical origin of this observed phenomenon to the formation of long wavelength charge order in the interfacial states in CrOCl and a band reconstruction in graphene subsequently. Our findings seem to open a new door of engineering the quantum Hall phase, and may shed light in the future manipulation of quantum electronic states via interfacial charge coupling, such as to construct novel topological superconductor, and to build quantum metrology standards.

\section{Methods}

$\textbf{Sample fabrications and characterizations.}$ The CrOCl/graphene/h-BN heterostructures were fabricated in ambient condition using the dry-transfer method, with the flakes exfoliated from high quality bulk crystals. CrOCl layers were etch patterned using an ion milling with Ar plasma, and dual gated samples are fabricated using standard e-beam lithography. A Bruker Dimension Icon atomic force microscope was used for thicknesses, morphology, and surface potential tests. The electrical performances of the devices were measured using a BlueFors LD250 at mK temperature, a Quantum Design PPMS system was used for temperature (3-300 K) and magnetic field ($\pm$14 T) scannings, and a probe station (Cascade Microtech Inc. EPS150) for room temperature electrical tests. 

$\textbf{Density functional theory calculations.}$ Density functional theory calculations.The first principles calculations based on density functional theory (DFT) are carried out with Vienna $ab$ $initio$ Simulation Package (VASP) with projector augmented wave method \cite{Kresse_PRB1996, Kresse_CMS1996}. The plane-wave energy cutoffis set to be 600 eV, and the crystal structure is fully relaxed until the residual forces on atoms are less than 0.01 eV/$\AA$. The generalized gradient approximation (GGA) by Perdew, Burke, and Ernzerhof is taken as the exchange–correlation potential \cite{Perdew_PRL1996}. Since Cr is a transition metal element with localized $3d$ orbitals, the on-site Hubbard U = 2.7 eV parameter is used in the calculations. To properly include the effects of vertical electric fields, we focus on a bilayer CrOCl because monolayer is indifferent to electric fields. We have considered three magnetic configurations in the CrOCl, and the thickness of the vacuum region was set as 40 $\AA$ to avoid any artificial interactions. The``DFT+D2'' type of vdW correction has been adopted for bulk calculations to properly describe the interlayer interactions \cite{Ref52, Ref53}. The so-calledfully localized limit of the spin-polarized GGA+U functional is adopted as suggested by Liechtenstein and coworkers \cite{Liechtenstein_PRB1995, Czyzyk_PRB1994, Anisimov_PRB1993, Solovyev_PRB1994}. In the calculations of the commensurate supercell of bilayer-CrOCl/graphene heterostructure, a 2$\times$4 supercell for bilayer CrOCl, and a 3$\times$3$\sqrt{3}$ supercell for monolayer graphene have been adopted.

\section{DATA AVAILABILITY}
The data that support the findings of this study will be available at the open-access repository Zenodo with a doi link, when published.

\section{Code AVAILABILITY}
The codes used in theoretical simulations and calculations are available from the corresponding authors on reasonable request.

\section{ACKNOWLEDGEMENT}
This work is supported by the National Key R$\&$D Program of China (2019YFA0307800, 2017YFA0303400, 2018YFA0306900, 2020YFA0309601, and 2018YFA0306101), and is supported by the National Natural Science Foundation of China (NSFC) with Grants 11974357, U1932151, 52031014, 11974340, 12174257, and 51627801. This work is also supported by the Strategic Priority Research Program of the Chinese Academy of Sciences (Grant No. XDB28000000), the MOST of China (Grants No. 2018YFA0306101 and No. 2017YFA0303400), the Chinese Academy of Sciences (Grants No. QYZDJ-SSW-SYS001, and No. XDPB22). Growth of hexagonal boron nitride crystals was supported by the Elemental Strategy Initiative conducted by the MEXT, Japan ,Grant Number JPMXP0112101001, JSPS KAKENHI Grant Number JP20H00354 and A3 Foresight by JSPS.

\section{Author contributions}
Z.H. and Y.Y. conceived the experiment. Z.H., Y.Y., J.Z., J.L. and Z.Z. supervised the overall project. X.L. and J.L. constructed the low-energy effective model. W.L., S.Z., J.L., J.K., K.C. performed the ab initio calculations. Y.G. performed the electro-static model. Y.W., X.G., K.Y., and N.R. carried out device fabrications, Y.W., X.G., K.Y., and P.G. performed electrical transport measurements; J.M. and Y.J. performed STM studies; K.W. and T.T. provided high quality h-BN bulk crystals. P.G. and Y.Y. carried out synthesis of CrOCl single crystals. Y.W., B.D., J.Z., J.M.,J.L., K.C., Y.Y., and Z.H. analysed the data. The manuscript was written by Z.H., J.L., and Y.W., with discussion and inputs from all authors.

\section{ADDITIONAL INFORMATION}
Competing interests: The authors declare no competing interests.

%\bibliography{arXiv}
%\bibliography{Ref_V1}
%\bibliographystyle{naturemag}

\end{document}